\documentclass[11pt]{amsart}
\usepackage{amsfonts}
\usepackage{amssymb}
\usepackage{amsmath}
\usepackage{amscd}
\begin{document}
\setlength{\oddsidemargin}{0cm} \setlength{\evensidemargin}{0cm}

\theoremstyle{plain}

\newtheorem{prop}{Proposition}
\newtheorem{theo}{Theorem}
\newtheorem{lemma}{Lemma}
\newtheorem{coro}{Corollary}
\newtheorem{defi}{Definition}
\newtheorem{exam}{Example}
\newtheorem{rem}{Remark}

\newcommand{\ba}{\begin{array}}
\newcommand{\ea}{\end{array}}
\newcommand{\bt}{\begin{tabular}}
\newcommand{\et}{\end{tabular}}
\newcommand{\btb}{\begin{table}}
\newcommand{\etb}{\end{table}}
\newcommand{\bc}{\begin{center}}
\newcommand{\ec}{\end{center}}
\newcommand{\bea}{\begin{eqnarray}}
\newcommand{\eea}{\end{eqnarray}}
\newcommand{\Bea}{\begin{eqnarray*}}
\newcommand{\Eea}{\end{eqnarray*}}
\newcommand{\beq}{\begin{equation}}
\newcommand{\eeq}{\end{equation}}
\newcommand{\Beq}{\begin{equation*}}
\newcommand{\Eeq}{\end{equation*}}

\def\blue {\color{blue}}
\def\red {\color{red}}

\def\vep{\varepsilon}
\def\vn{\varepsilon}
\def\ot{\otimes}
\def\om{\omega}
\def\q{\boldsymbol{q}}
\def\bv{\boldsymbol{v}}
\def\bc{\boldsymbol{c}}
\def\lan{\langle}
\def\ran{\rangle}
\def\al{\alpha}
\def\th{\theta}
\def\be{\beta}
\def\De{\Delta}
\def\ga{\gamma}
\def\Ga{\Gamma}
\def\Om{\Omega}
\def\si{\sigma}
\def\tu{\tilde{u}}
\def\ep{\epsilon}
\def\de{\delta}
\def\pa{\partial}
\def\La{\Lambda}
\def\la{\lambda}
\def\bi{\binom}
\def\lra{\longrightarrow}
\def\lmto{\longmapsto}
\def\ns{\mathcal{NS}}

\newcommand{\cL}{{\mathcal{L}}}
\newcommand{\E}{{\mathcal{E}}}
\newcommand{\M}{{\mathcal M}}
\newcommand{\R}{{\mathcal R}}
\newcommand{\N}{{\mathcal N}}
\newcommand{\C}{{\mathcal C}}
\newcommand{\D}{{\mathcal D}}
\newcommand{\B}{{\mathcal B}}
\newcommand{\F}{{\mathcal F}}
\newcommand{\Z}{{\mathcal Z}}
\newcommand{\K}{{\mathcal K}}
\newcommand{\cI}{{\mathcal I}}
\newcommand{\J}{{\mathcal J}}
\newcommand{\A}{{\mathcal A}}
\newcommand{\G}{{\mathcal G}}
\newcommand{\U}{{\mathcal U}}
\newcommand{\CO}{{\mathcal O}}
\def\I{{\mathcal{I} }}

\def\bN{{\mathbb N}}
\def\bZ{{\mathbb Z}}
\def\bQ{{\mathbb Q}}
\def\bR{{\mathbb R}}
\def\bT{{\mathbb T}}
\def\bF{{\mathbb F}}
\def\bK{{\mathbb K}}
\def\bC{{\mathbb C}}

\def\sA{{\mathscr A}}
\def\P{{\mathcal P}}
\def\sB{{\mathscr B}}
\def\C{{\mathscr C}}
\def\sL{{\mathscr L}}

\def\mh{\mathfrak{h}}
\def\b{\mathfrak{b}}
\def\n{\mathfrak{n}}
\def\mg{\mathfrak{g}}
\def\H{{\mathscr H}}

\def\Inv{\mbox{\rm Inv}}
\def\Res{\mbox{\rm Res}}
\def\Der{\mbox{\rm Der}}
\def\Diag{\mbox{\rm Diag}}
\def\rank{\mbox{\rm rank}}
\def\Ob{\mbox{\rm Ob}}
\def\ad{\mbox{\rm ad}}
\def\Hom{\mbox{\rm Hom}}
\def\op{\mbox{\scriptsize op}}
\def\ext{\mbox{\rm Ext}\,}
\def\Ker{\mbox{\rm Ker}\,}
\def\udim{{\mathbf {\dim}\,}}
\def\mo{\mbox{\rm mod}\,}
\def\mx{\mbox{\rm max}}
\def\tr{\mbox{\rm tr}\,}
\def\rad{\mbox{\rm rad}\,}
\def\top{\mbox{\rm top}\,}
\def\rep{\mbox{\rm Rep}\,}
\def\End{\text{\rm End}}
\def\Im{\text{\rm Im}}
\def\id{\text{\rm id}}
\def\wt{\text{\rm wt}}
\def\ss{\scriptscriptstyle}
\def\scr{\scriptstyle}
\def\e{\mbox{\rm e}}
\def\uf{\mbox{\bF}}
\def\f{{\mathbf {\uf}}}

\baselineskip=20pt

\title{An operator approach to the rational solutions of the classical Yang-Baxter
equation}

\author{Qiang Zhang}
\address{Chern Institute of Mathematics \& LPMC, Nankai University,
Tianjin 300071, P.R. China} \email{qiangifang@mail.nankai.edu.cn}

\author{Chengming Bai
$^*$ }

\address{Chern Institute of Mathematics \& LPMC, Nankai University,
Tianjin 300071, P.R. China} \email{baicm@nankai.edu.cn}
\thanks{$^*$ Corresponding author}

\def\shorttitle{An Operator approach to the rational solutions of the CYBE}

\subjclass[2000]{81R, 17B}


\begin{abstract}

Motivated by the study of the operator forms of the constant
classical Yang-Baxter equation given by Semonov-Tian-Shansky,
Kupershmidt and the others, we try to construct the rational
solutions of the classical Yang-Baxter equation with parameters by
certain linear operators. The fact that the rational solutions of
the CYBE for the simple complex Lie algebras can be interpreted in
term of certain linear operators motivates us to give the notion of
$\mathcal O$-operators such that these linear operators are the
$\mathcal O$-operators associated to the adjoint representations.
Such a study can be generalized to the Lie algebras with
nondegenerate symmetric invariant bilinear forms. Furthermore we
give a construction of a rational solution of the CYBE from an
$\mathcal O$-operator associated to the coadjoint representation and
an arbitrary representation with a trivial product in the
representation space respectively.

\end{abstract}

\maketitle

\section{Introduction}

The classical Yang-Baxter equation (CYBE) first arose in the study
of the inverse scattering theory (see \cite{FT1}, \cite{FT2}) and
has played an important role in the study of the classical
integrable systems (\cite{Be}, \cite{GD}, \cite{STS2}, \cite{STS3},
\cite{STS4}, \cite{STS1}, \cite{Sk}, etc.). There are some close
relations between it and many branches of mathematical physics and
pure mathematics, like symplectic geometry, quantum groups, quantum
field theory and so on (see \cite{CP} and the references therein).

The classical Yang-Baxter equation with spectral parameters is given
as
$$[[r,r]]=[r_{12}(u_1,u_2),r_{13}(u_1,u_3)]+[r_{12}(u_1,u_2),r_{23}(u_2,u_3)]+[r_{13}(u_1,u_3),r_{23}(u_2,u_3)]=0,\eqno (1.1)$$
where $r$ is a function $r:
\mathbb{F}\otimes\mathbb{F}\rightarrow\frak g\otimes\frak g$ with
$\frak g$ being a Lie algebra over a field $\mathbb F$ and the
notations $r_{ij}$ are given as follows. For any $r=\sum\limits_i
a_i\otimes b_i\in \frak g\otimes \frak g$, set
$$r_{12}=\sum\limits_i a_i\otimes b_i\otimes 1,\;\; r_{13}=\sum\limits_i
a_i\otimes1\otimes b_i,\;\;r_{23}=\sum\limits_i 1\otimes a_i\otimes
b_i,\eqno (1.2)$$  and the commutation relations in (1.1) are given
in the universal enveloping algebra $U(\frak g)$ of the Lie algebra
$\frak g$.

Most of the study on the classical Yang-Baxter equation (1.1) is
concentrated on the following cases (\cite{BD},
\cite{BD1},\cite{D1}, \cite{St3}, etc.): $\frak g$ is taken as a
finite-dimensional simple Lie algebra over the complex number field
$\mathbb C$ and $r$ is nondegenerate which depends on a single
parameter. That is, $r$ satisfies
$$r(u_1,u_2)=r(u_1-u_2),\;\;\eqno (1.3)$$
and there is no proper subalgebra $\frak h$ of $\frak g$ such that
$r(u)\in \frak h\otimes\frak h$.

According to Belavin and Drinfeld (\cite{BD},\cite{BD1}), the
nondegenerate solutions of the classical Yang-Baxter equation (1.1)
depending on a single parameter for the simple complex Lie algebras
are divided into three cases: trigonometric, elliptic and rational.
In this paper, we pay our main attention to the rational solutions
$r$ with exactly one pole. In fact, a general form of a rational
solution $r$ of the CYBE can be written as (\cite{BD},\cite{BD1},
\cite{D1},
 \cite{St3},\cite{St1},\cite{St2})
$$r(u_1,u_2)=\frac{t}{u_1-u_2}+r_0(u_1,u_2),\eqno (1.4)$$ where $t$
is the Casimir element of $\frak g$ and $r_0$ is a polynomial in
$\frak g[u_1]\otimes \frak g[u_2]$. However, it is not easy to get
an explicit expression of $r_0$ from the equation (1.4).
 Moreover, it is also difficult to extend the
study from the simple complex Lie algebras to the other Lie algebras.

On the other hand, for any $r\in \frak g\otimes \frak g$, $r$ can be
expressed by a matrix under a basis. So it is natural to consider
the conditions satisfied by the linear maps corresponding to the
matrices (classical $r$-matrices) satisfying the CYBE. For the
constant solutions of the CYBE, Semonov-Tian-Shansky (\cite{STS2})
first gave an operator form of the CYBE  as a linear map $R:\frak
g\rightarrow \frak g$ satisfying
$$[R(x), R(y)]=R([R(x),y]+[x,R(y)]),\;\;\forall x,y\in \frak g.\eqno (1.5)$$ It is equivalent to the tensor form of the CYBE when the
following two conditions are satisfied: (a) there exists a
nondegenerate symmetric invariant bilinear form on $\frak g$ and (b)
$r$ is skew-symmetric. Note that equation (1.5) is exactly the
Rota-Baxter relation of weight-zero in the version of Lie algebras
(\cite{Ag}, \cite{E}, \cite{EGK}), whereas the Rota-Baxter relations
were introduced to generalizes the integration-by-parts formula
(\cite{Bax},\cite{Rot1},\cite{Rot2}) and then (the versions of
associative algebras) play important roles in many fields in
mathematics and mathematical physics (cf. \cite{AB} and the
references therein).

Furthermore, Kupershmidt (\cite{Ku}) replaced the above condition
(a) by letting $r$ be a linear map from $\frak g^*$ to $\frak g$ and
when $r$ is skew-symmetric, the tensor form of the CYBE is
equivalent to such a linear map $r$ satisfying
$$[r(a^*),r(b^*)]=r({\rm ad}^* r(a^*)(b^*)-{\rm
ad}^*r(b^*)(a^*)),\;\forall a^*,b^*\in {\frak g}^*,\eqno (1.6)$$
where ${\frak g}^*$ is the dual space of ${\frak g}$ and ${\rm
ad}^*$ is the dual representation of the adjoint representation
(coadjoint representation) of the Lie algebra $\frak g$. Moreover,
Kupershmidt generalized the above ${\rm ad}^*$ to be an arbitrary
representation $\rho:\frak g\rightarrow gl(V)$ of $\frak g$, that
is, a linear map $T:V\rightarrow \frak g$ satisfying
$$[T(u), T(v)]=T(\rho(T(u))v-\rho(T(v))u),\forall u,v\in V,\eqno (1.7)$$
which was regarded as a natural generalization of the CYBE. Such an
operator $T$ is called an ${\mathcal O}$-operator associated to
$\rho$ by Kupershmidt (\cite{Ku}). It was also mentioned in
\cite{Bo}. Moreover, such an $\mathcal O$-operator indeed gives a
constant solution of the CYBE in a larger Lie algebra (\cite{Bai}).

Then it is natural to consider how to extend such an idea to study
the rational solutions of the CYBE (1.1), which is the main aim in
our paper. We would like to point out that this study is not a
simple generalization since it is quite different with the study
 of the constant solutions  (see the discussion in
Section 5), although the idea is quite similar to the study in
\cite{Bai}. On the other hand, Xu also considered to use the
operator form to study the CYBE (1.1) in \cite{X} (even he extended
his study to any nonassociative algebra). We would like to point out
that although the ideas are quite similar (which both are in fact
motivated by the study of Semonov-Tian-Shansky (\cite{STS2})), they
are two different approaches. One of the main differences is that
Xu's approach is the direct generalization of equation (1.5) with a
similar form (thus the existence of nondegenerate associative
symmetric bilinear form and the skew-symmetry is necessary for his
study on the general nonassociative algebras including Lie algebras)
and he focused on the trigonometric solutions with a similar form on
certain more general algebras, whereas our approach are essentially
the generalizations of equations (1.5)-(1.7) with ceratin
``modified" forms for a general Lie algebra without many additional
constraints and we paid our main attention to the rational solutions
with the form (1.4). More comparisons between the two approaches are
given in the following sections.

The paper is organized as follows. In Section 2, we interpret the
rational solutions of the CYBE for the simple complex Lie algebras
in term of certain linear operators which motivates us to give the
notion of $\mathcal O$-operators such that these linear operators
are the $\mathcal O$-operators associated to the adjoint
representations. Such a study can be generalized to the Lie algebras
with nondegenerate symmetric invariant bilinear forms. In Section 3,
we generalize the Casimir element appearing in the rational
solutions of the CYBE in Section 2 to a symmetric invariant tensor
under the action of the adjoint representation, which gives a
construction of a rational solution of the CYBE from an ${\mathcal
O}$-operator associated to the coadjoint representation. In Section
4, we give a construction of a rational solution of the CYBE from an
${\mathcal O}$-operator associated to an arbitrary representation
with a trivial product in the representation space. In Section 5, we
give some conclusions and discussion.

\section{An $\mathcal {O}-$operator associated to a rational solution
of the CYBE for a Lie algebra with a nondegenerate symmetric
invariant bilinear form}

Let $\frak g$ be a Lie algebra. Let $\sigma:\frak g\otimes \frak
g\rightarrow \frak g\otimes \frak g$ be the exchanging operator
satisfying $\sigma(x\otimes y)=y\otimes x$ for any $x,y\in \frak g$.
For any $r=\sum\limits_ia_i\otimes b_i$, we set
$$r^{21}=\sigma(r)=\sum\limits_i b_i\otimes a_i.\eqno(2.1)$$
A bilinear form ${\mathcal B}$ on $\frak g$ is invariant if $\mathcal B$ satisfies
$$\mathcal B ([x,y],z)=\mathcal B(x, [y,z]),\;\;\;\forall x,y,z\in \frak g.\eqno (2.2)$$

We begin our study from the case of $\frak g$ being a simple complex Lie algebra. Let $k(\;,\;)$ be the
Killing form on $\frak g$ which is the unique nondegenerate symmetric invariant bilinear form on $\frak g$ up to a scalar multiplication.
 Let $r$ be a nondegenerate rational solution of the CYBE (1.1). In addition, $r$ usually satisfies the
unitary condition:
$$r(u_1,u_2)+r^{21}(u_2,u_1)=0.\eqno (2.3)$$
As in the Introduction, a general form of $r$ is given as
$$r(u_1,u_2)=\frac{t}{u_1-u_2}+r_0(u_1,u_2),\eqno (2.4)$$
where $t=\sum\limits_ie_i\otimes e_i$ is the Casimir element of
$\frak g$, \{${e_i}$\} is an orthonormal basis of $\frak g$
associated to the Killing form $k(\;,\;)$ and $r_0(u_1,u_2)\in \frak
g[u_1]\otimes \frak g [u_2]$. According to Stolin's study in
\cite{St3}, \cite{St1} and \cite{St2}, we can set
$$r_0(u_1,u_2)=\sum_{i=1}^{K}\sum_{p=0}^{M}\mu(e_iu_1^{-p-1})\otimes
e_iu_2^p,\eqno (2.5)$$ where $\mu $ is a linear operator from $\frak
g[u^{-1}]u^{-1}$ to $\frak g[u]$, $M,K\in\mathbb{N},$ and dim$\frak
g=K.$  Note that Stolin has proved that $\deg_{u_i} r_0\leq1$ when
$\frak g$ is the simple Lie algebra $sl(n)$. But it is not necessary
to consider this conclusion because the following study can be
generalized to some more general Lie algebras. On the other hand, in
\cite{X}, the operator form $r'(z)$ related to a solution $r(z)$  of
the CYBE (1.1) satisfying equation (1.3) is given by
$$r(z)=\sum_{i\in \Omega} r'(z)(e_i)\otimes e_i,\eqno (2.6)$$
where $\{e_i|i\in \Omega\}$ is a basis of $\frak g$, $r(z)$ is a
function with domain $\frak D\subset \mathbb C$ and range $\frak
g\otimes \frak g$, $r'(z)\in {\rm End}(\frak g)$. Comparing
equations (2.5) and (2.6), we know that the domain of the linear
operator $\mu$ in equation (2.5) is $\frak g[u^{-1}]u^{-1}$ (later
we will extend it to be the whole algebra $\frak g[u,u^{-1}]$),
whereas the linear operator $r'(z)$ appearing in equation (2.6) can
be regarded as a family of the linear transformations on $\frak g$
with the parameter $z$. In fact, the latter $r'(z)$ gives a kind of
trigonometric solutions from an identity on $e^z$ (\cite{X}).

 Substituting the form (2.5) into the CYBE (1.1), we have
\begin{eqnarray*}
\hspace{1cm}[[r,r]]&=&\sum_{i,j=1}^K\sum_{p,q=0}^{M}[\mu(e_iu_1^{-p-1}), \mu(e_ju_1^{-q-1})]\otimes e_iu_2^p\otimes e_ju_3^q\\
      &+&\sum_{i,j=1}^K\sum_{p,q=0}^{M}\mu(e_iu_1^{-p-1})\otimes[e_iu_2^p, \mu(e_ju_2^{-q-1})]\otimes
      e_ju_3^q\\
      &-&\sum_{i,j=1}^K\sum_{p,q=0}^{M}\mu(e_iu_1^{-p-1})\otimes e_ju_2^q\otimes [e_iu_3^p, \mu(e_ju_3^{-q-1})]\\
      &+&\sum_{i,s=1}^K\sum_{p=0}^{M}\mu(e_iu_1^{-p-1})[e_s, \frac{\mu(e_iu_1^{-p-1})-\mu(e_iu_2^{-p-1})}{u_1-u_2}]\otimes e_s\otimes
      e_iu_3^p\\
      &+&\sum_{i,s=1}^K\sum_{p=0}^{M}\frac{[\mu(e_iu_1^{-p-1})-\mu(e_iu_3^{-p-1}),
e_s]}{u_1-u_3}\otimes e_iu_2^p\otimes
      e_s\\
      &+&\sum_{i,s=1}^K\sum_{p=0}^{M}\mu(e_iu_1^{-p-1})\otimes [e_i, e_s]\otimes
      e_s\frac{u_2^p-u_3^p}{u_2-u_3}.\hspace{5.3cm} (2.7)
\end{eqnarray*}
Since $r$ satisfies the
unitary condition (2.3), we know that
$$\sum_{j,q}\mu(e_ju_1^{-q-1})\otimes e_ju_2^q+\sum_{j,q} e_ju_1^q\otimes \mu(e_ju_2^{-q-1})=0;\eqno (2.8)$$
that is,
$$\sum_{j,q}\mu(e_ju_1^{-q-1})\otimes e_ju_2^q \;\;{\rm can}\;\;{\rm be}\;\;{\rm replaced}\;\;{\rm by}\;\;
-\sum_{j,q} e_ju_1^q\otimes \mu(e_ju_2^{-q-1}).\eqno (2.9)$$
Furthermore, due to the unitary condition (2.3) again, we know that
$\deg(\mu (e_iu^{-p-1}))\leq M$. Hence we can let
$$\mu (e_iu^{-p-1})=\sum_{l=0}^M\alpha_l(e_iu^{-p-1})u^l,\eqno (2.10)$$
where $\alpha_l$ is a linear operator from $\frak g[u^{-1}]u^{-1}$
to $\frak g$, $l=0,1,\cdots, M$. Since $r$ is a rational function
and $r_0$ is a polynomial, $\mu$ can be defined on the whole $\frak
g[u^{-1}]u^{-1}$ by the zero-extension. Set
$$\alpha_l\equiv0,\;\;{\rm when}\;\; l>M.\eqno(2.11)$$

We divide the right hand side of the equation (2.7) into four parts
$$[[r,r]]=\sum_{i,j=1}^K\sum_{p,q=0}^{M}[\mu(e_iu_1^{-p-1}), \mu(e_ju_1^{-q-1})]\otimes e_iu_2^p\otimes e_ju_3^q+(A)+(B)+(C),\eqno(2.12)$$
where \begin{eqnarray*}
\hspace{2cm}(A)&=&\sum_{i,j=1}^K\sum_{p,q=0}^{M}\mu(e_iu_1^{-p-1})\otimes[e_iu_2^p, \mu(e_ju_2^{-q-1})]\otimes e_ju_3^q\\
& &+\sum_{i,s=1}^K\sum_{p=0}^{M}[e_s, \frac{\mu(e_iu_1^{-p-1})-\mu(e_iu_2^{-p-1})}{u_1-u_2}]\otimes e_s\otimes e_iu_3^p;\hspace{2.5cm} (2.13)\\
(B)&=&-\sum_{i,j=1}^K\sum_{p,q=0}^{M}\mu(e_iu_1^{-p-1})\otimes e_ju_2^q\otimes [e_iu_3^p, \mu(e_ju_3^{-q-1})]\\
& &+\sum_{i,s=1}^K\sum_{p=0}^{M}[\frac{\mu(e_iu_1^{-p-1})-\mu(e_iu_3^{-p-1})}{u_1-u_3}, e_s]\otimes e_iu_2^p\otimes e_s;\hspace{2.5cm} (2.14)\\
(C)&=&\sum_{i,s=1}^K\sum_{p=0}^{M}\mu(e_iu_1^{-p-1})\otimes [e_i,
e_s]\otimes e_s\frac{u_2^p-u_3^p}{u_2-u_3}.\hspace{4.3cm} (2.15)
\end{eqnarray*}
It is easy to know that
$${\mathcal B}(f,g)=-{\rm Res}_{u=0}k(f,g),\;\;\forall\;\;f,g\in \frak g[u,u^{-1}]\eqno (2.16)$$
is an invariant bilinear form on the Lie algebra $\frak g[u,u^{-1}]$. Hence for any $f\in \frak g[u]$, we know that
$$f=\sum_{i=1}^K\sum_{p=0}^\infty \mathcal B(f,e_iu^{-p-1})e_iu^p,\eqno (2.17)$$
where there are always finite terms not zero in the above equation.
Extend the linear operator $\mu$ from $\frak g[u^{-1}]u^{-1}$ to
$\frak g[u,u^{-1}]$ by ($id$ is the identity
operator)
$$\mu\mid_{\frak g[u]}=-id.\eqno (2.18)$$ Therefore
\begin{eqnarray*}
\hspace{1cm}(A)
&=&\sum_{i,j,k=1}^K\sum_{q,n=0}^{M}\sum_{p=0}^{2M}\mu(e_iu_1^{-n-1})\otimes
\mathcal B(
              [e_iu^n,\mu(e_ju^{-q-1})], e_ku^{-p-1})
              e_ku_2^p\otimes e_ju_3^q\\
              &&+\sum_{i,s=1}^K\sum_{p=0}^{M}[e_s, \sum_{l=0}^{M}\sum_{q=0}^l\alpha_{l+1}(e_iu_1^{-p-1})u_1^{l-q}]\otimes
              e_su_2^q\otimes e_iu_3^p\\
                           &=&\sum_{i,j,k=1}^K\sum_{q,n=0}^{M}\sum_{p=0}^{2M} \mathcal B ([e_iu^n,\mu(e_ku^{-q-1})],
              e_ju^{-p-1}) \mu(e_iu_1^{-n-1})\otimes
              e_ju_2^p\otimes e_ku_3^q\\
              &&+\sum_{i,j=1}^K\sum_{p=0}^{M}\sum_{l=0}^{M}\sum_{q=0}^l[e_iu_1^{l-q}, \alpha_{l+1}(e_ju_1^{-p-1})]\otimes
              e_iu_2^q\otimes e_ju_3^p\\
              \end{eqnarray*}
\begin{eqnarray*}
             \hspace{1.7cm} &=&\sum_{i,j,k=1}^K\sum_{q,n=0}^{M}\sum_{p=0}^{2M} \mathcal B(e_iu^n,[\mu(e_ku^{-q-1}),e_ju^{-p-1}])
             \mu(e_iu_1^{-n-1})\otimes
              e_ju_2^p\otimes e_ku_3^q\\
              &&+\sum_{i,j=1}^K\sum_{q,p=0}^{M}\sum_{l=p}^{M-1}[e_iu_1^{l-p}, \alpha_{l+1}(e_ju_1^{-q-1})]\otimes
              e_iu_2^p\otimes e_ju_3^q\\
           &=&\sum_{i,j=1}^K\sum_{q=0}^{M}\sum_{p=0}^{2M}\mu[\sum_{l=0}^p\alpha_{l}(e_ju^{-q-1})u_1^l,
              e_iu_1^{-p-1}]\otimes e_iu_2^p\otimes e_ju_3^q\\
              &&-\sum_{i,j=1}^K\sum_{p,q=0}^{M}\sum_{l=p+1}^{M}[\alpha_{l}(e_ju_1^{-q-1})u_1^{l},
              e_iu_1^{-p-1}]\otimes e_iu_2^p\otimes e_ju_3^q\\
&=&\sum_{i,j=1}^K\sum_{q=0}^{M}\sum_{p=M+1}^{2M}\mu[\mu(e_ju_1^{-q-1}),
e_iu_1^{-p-1}]\otimes
              e_iu_2^p\otimes e_ju_3^q\\
&&+\sum_{i,j=1}^K\sum_{q=0}^{M}\sum_{p=0}^{M}\mu[\sum_{l=0}^p\alpha_{l}(e_ju^{-q-1})u_1^l,
              e_iu_1^{-p-1}]\otimes e_iu_2^p\otimes e_ju_3^q\\
&&-\sum_{i,j=1}^K\sum_{p,q=0}^{M}\sum_{l=p+1}^{M}[\alpha_{l}(e_ju_1^{-q-1})u_1^{l},
              e_iu_1^{-p-1}]\otimes e_iu_2^p\otimes e_ju_3^q\\
 &=&\sum_{i,j=1}^K\sum_{q=0}^{M}\sum_{p=0}^{2M}\mu[\mu(e_ju_1^{-q-1}),
e_iu_1^{-p-1}]\otimes
              e_iu_2^p\otimes e_ju_3^q.\hspace{4cm} (2.19)
\end{eqnarray*}
Note that for convenience,  all the degree parameters can be taken
from 0 to $2M$ (here and in the following sections), that is,
$$(A)=\sum_{i,j=1}^K\sum_{p,q=0}^{2M}\mu[\mu(e_ju_1^{-q-1}),
e_iu_1^{-p-1}]\otimes
              e_iu_2^p\otimes e_ju_3^q.\eqno (2.20)$$
Similarly,
$$(B)=-\sum_{i,j=1}^K\sum_{p,q=0}^{2M}\mu[\mu(e_iu_1^{-p-1}),
e_ju_1^{-q-1}]\otimes
             e_iu_2^p\otimes e_ju_3^q.\eqno (2.21)$$
Set $[e_i,e_s]=C_{is}^ke_k$, where (and in the following) the
repeated (up and down) indices mean summation. Then
$C_{is}^k=C_{sk}^i$ and we have
\begin{eqnarray*}
\hspace{3cm}(C)&=&\sum_{i,s=1}^K\sum_{p,q=0}^{2M}\mu(e_iu_1^{-p-q-2})\otimes
[e_i, e_s]u_2^p\otimes
             e_su_3^q\\
             &=&\sum_{i,s,k=1}^K\sum_{p,q=0}^{2M}\mu(C_{is}^ke_iu_1^{-p-q-2})\otimes e_ku_2^p\otimes e_su_3^q\\
             &=&\sum_{s,k=1}^K\sum_{p,q=0}^{2M}\mu[e_su_1^{-q-1},e_ku_1^{-p-1}]\otimes e_ku_2^p\otimes e_su_3^q.\hspace{2.95cm} (2.22)
\end{eqnarray*}
Therefore, we have the following conclusion.

{\bf Theorem 1}\quad Let $\frak g$ be a complex simple Lie algebra.
Let $r$ be given by the equation (2.4) satisfying the unitary
condition (2.3). Then $r$ is a nondegenerate solution of the CYBE
for $\frak g$ if and only if the linear operator $\mu: \frak
g[u,u^{-1}]\rightarrow \frak g [u]\subset \frak g[u,u^{-1}]$
defining the polynomial $r_0$ by the equation (2.5) satisfies
$\mu|_{\frak g[u]}=-id$ and
$$[\mu(f),\mu(g)]=\mu[\mu(f),g]+\mu[f,\mu(g)]+\mu[f, g],\;\;\forall f,
g\in\frak g[u,u^{-1}],\eqno (2.23)$$
$$\sum_{i=1}^{K}\sum_{p=0}^{M}(\mu(e_iu_1^{-p-1})\otimes
e_iu_2^p+e_iu_1^p\otimes \mu(e_iu_2^{-p-1}))=0.\eqno(2.24)$$

In fact, the above conclusion can be implied by an (equivalent)
result in \cite{St1} given by Stolin with a different approach (see
Theorem 1.1 and its proof in \cite{St1}) as follows. When $\frak g$
is a simple Lie algebra, as a key point of his study, Stolin  proved
that there is a natural one-to-one correspondence between the
rational solutions of CYBE in $\frak g$ and the subspaces $W\subset
\frak g((u^{-1}))$ (where $\frak g((u^{-1}))=\{
\sum\limits_{i=-\infty}^m x_iu^i|x_i\in\frak g,\;\;{\rm certain}\;\;
m\in \mathbb N\}$) such that

(a) $W$ is a subalgebra in $\frak g((u^{-1}))$ such that $W\supset
 u^{-N} g[[u^{-1}]]$ for some $N>0;$

(b) $W\otimes \frak g[u]=\frak g((u^{-1}));$

(c) $W$ is a Lagrangian subspace with respect to the bilinear form
$\mathcal B'$ of $\frak g((u^{-1}))$ given by
 $$\mathcal B'(f, g)=-{\rm Res}_{u=0}\;\mathcal B(f , g ),\;\;\forall f, g\in
 \frak g((u^{-1})).\eqno (2.25)$$
It is straightforward to prove that the linear operator $\mu$ given
by equation (2.23) (the domain can be extended to $ u^{-1}\frak
g[[u^{-1}]]$ by zero-extension) is exactly the operator satisfying
$$\mathcal B'([f+\mu (f),g+\mu(g)],h+\mu(h))=0,\;\; \forall f,g,h\in u^{-1}\frak
g[[u^{-1}]],\eqno (2.26)$$ which was given by Stolin to decide the
corresponding subspace $W\subset \frak g((u^{-1}))$ satisfying the
above three conditions by
$$W=(1+\mu)u^{-1}\frak g[[u^{-1}]].\eqno (2.27)$$
Note that the study of Stolin on the correspondence between the
rational solutions of CYBE and the subspaces $W\subset \frak
g((u^{-1}))$ is valid only for $\frak g$ being a complex simple Lie
algebra.

We call a linear operator $\mu:\frak g[u,u^{-1}]\rightarrow \frak
g[u,u^{-1}]$ satisfying the equation (2.23) an $\mathcal
{O}$-operator. In fact, in the next section, we will give an exact
definition of an $\mathcal O$-operator associated to any
representation which the notion is due to its similarity with the
notion $\mathcal O$-operator (1.7) for the constant CYBE given by
Kupershmidt (\cite{Ku}). In this sense, equation (2.23) just gives
an $\mathcal O$-operator  associated to the adjoint representation.
On the other hand, note that equation (2.23) is exactly the
Rota-Baxter relation of weight $-1$ in the version of Lie algebras
(\cite{E}, \cite{EGK}). Obviously, such a notion of an $\mathcal
O$-operator (the equation (2.23)) can be defined for any Lie
algebra. Furthermore, note that in the above study, the
nondegenerate symmetric invariant bilinear form (the Killing form)
on the Lie algebra $\frak g$ plays an essential role. So by a
similar study, we can extend Theorem 1 as follows (which is a new
conclusion to our knowledge).

{\bf Theorem 2}\quad Let $\frak g$ be a Lie algebra with a
nondegenerate symmetric invariant bilinear form. Let $\{e_i\}$ be an
orthonormal basis of $\frak g$ associated to the bilinear form and
$t=\sum\limits_ie_i\otimes e_i$. Let $r$ be given by the equation
(2.4) satisfying the unitary condition (2.3). Then $r$ is a
nondegenerate solution of the CYBE for $\frak g$ if and only if the
linear operator $\mu: \frak g[u,u^{-1}]\rightarrow \frak g[u]\subset
\frak g[u,u^{-1}]$ defining the polynomial $r_0$ by the equation
(2.5) is an $\mathcal O$-operator (that is, the equation (2.23)
holds) satisfying $\mu|_{\frak g[u]}=-id$ and the equation (2.24).

Note that when the Lie algebra $\frak g$ is simple, we can get all
the nondegenerate rational solutions satisfying unitary condition
(2.3) from the $\mathcal O$-operators as we have interpreted after
Theorem 1 (\cite{BD}, \cite{BD1},\cite{D1},\cite{St3},
\cite{St1},\cite{St2}). But it may fail for the general case. In
fact, the ${\mathcal O}$-operators for the Lie algebras with
nondegenerate symmetric invariant bilinear forms are only
``sufficient", that is, they can only give a kind of the rational
solutions of the CYBE (maybe not all!).

Furthermore, for a Lie algebra with a nondegenerate symmetric
invariant bilinear form,  Xu in \cite{X} gave another kind of
operator form (see equation (2.6) for the notations)
$$[r'(z_1+z_2)(x),
r'(z_2)(y)]=r'(z_1+z_2)[x,r'(-z_1)(y)]+r'(z_2)[r'(z_1)(x),y],\;\;
\forall x,y\in \frak g,\eqno (2.28)$$ which is equivalent to the
CYBE (1.1) under certain more conditions.
 Obviously, it is quite
different with Theorem 2 (also see the comparison the differences
between equations (2.5) and (2.6) given at the beginning of this
section).

On the other hand, one may think that the above study on the
rational solutions of the CYBE by introducing the notion of an
$\mathcal O$-operator is not very effective since merely the terms
of the polynomial part $r_0$ of $r$ have been concerned. This
weakness is rather evident when $\frak g$ is taken as a general Lie
algebra with a nondegenerate symmetric invariant bilinear form
because there probably exist other forms of the rational solutions
of the CYBE. In fact, it would not be difficult to give a definition
which covers the whole $r$ by considering how to extend the terms
with certain poles (see \cite{LL}). However, the corresponding
operator product expansion would be very complicated and it would
not be easy to give a further study explicitly since one might be
entangled with paying more attention to the parameter $u$.

At the end of this section, we give a special example of
constructing a rational solution of the CYBE from an
$\mathcal{O}$-operator for the classical double of a Lie bialgebra.
Recall that a Lie bialgebra structure on a Lie algebra $\frak g$ is
a skew-symmetric linear map $\delta_{\frak g}: {\frak g}\rightarrow
{\frak g}\otimes {\frak g}$ such that $\delta_{\frak g}^*:{\frak
g}^*\otimes {\frak g}^*\rightarrow {\frak g}^*$ is a Lie bracket on
${\frak g}^*$ and
$$\delta ([x,y])=[x\otimes 1+1\otimes x,\delta (y)]-[y\otimes
1+1\otimes y, \delta(x)],\;\;\forall x,y\in \frak g.\eqno (2.29)$$
It is equivalent to a Manin triple $(\frak g, \frak g^*, \mathcal
{B})$, that is, there is a Lie algebra structure on a direct sum
$\frak g\oplus\frak g^*$ of the underlying vector spaces of $\frak
g$ and $\frak g^*$ such that $\frak g$ and $\frak g^*$ are
subalgebras and the natural symmetric bilinear form on ${\frak
g}\oplus {\frak g}^*:$
$${\mathcal B}(x+a^*,y+b^*)=\langle a^*,y\rangle  +\langle x,b^*\rangle  ,\;\;
\forall x,y\in {\frak g},a^*,b^*\in {\frak g}^*,\eqno (2.30)$$ is
invariant, where $\langle, \rangle$ is the ordinary pair between
$\frak g$ and $\frak g^*$. The Lie algebra $\frak g\oplus \frak g^*$
with the bilinear form (2.30) is still a Lie bialgebra which is
called a classical double of the Lie bialgebra $(\frak
g,\delta_{\frak g})$ (\cite{CP}). Let $\{e_1,\cdots,e_K\}$ be a
basis of $\frak g$ and $\{e^*_1,\cdots, e^*_K\}$ be its dual basis.
Set
 $$[e_i,e_j]=C_{ij}^ke_k,\;\;[e^*_i,e^*_j]=\Gamma^k_{ij}e^*_k.\eqno
 (2.31)$$
Then the Lie algebraic structure on the classical double $\frak
g\oplus\frak g^*$ satisfies
$$[e^*_i,e_j]=\sum_k(C_{jk}^ie^*_k-\Gamma^j_{ik}e_k).\eqno (2.32)$$
and the bilinear form (2.30) is invariant. It is obvious that the
bilinear form on $(\frak g^*\oplus\frak g)[u,u^{-1}]$ given by
 $$\mathcal B'(f, g)=-{\rm Res}_{u=0}\;\mathcal B(f , g ),\;\;\forall f, g\in
 (\frak g\oplus \frak g^*)[u,u^{-1}]\eqno (2.33)$$  is invariant, too. Furthermore,
 let $\mu:(\frak g\oplus\frak g^*)[u,u^{-1}]\rightarrow
(\frak g\oplus\frak g^*)[u]$ be a linear operator satisfying the following conditions:

  (1) $\mu$ is an $\mathcal {O}$-operator $\mu$ on the Lie algebra
 $(\frak g^*\oplus\frak g)[u,u^{-1}]$, that is, $\mu$ satisfies
$$[\mu(f),\mu(g)]=\mu[\mu(f),g]+\mu[f,\mu(g)]+\mu[f, g] \qquad \forall
f,g\in(\frak g\oplus\frak g^*)[u^{-1},u];\eqno (2.34)$$

 (2) $\mu|_{(\frak g\oplus\frak g^*)[u]}=-id$;

 (3) There exists an $L\in\mathbb{N}$ such that $\mu(xu^{-n-1})=0$ for any $n>L$ and $x\in\frak g\oplus
     \frak g^*$. Moreover,
$$\sum_{i=1}^K\sum_{n=0}^L\{[\mu(e^*_iu_1^{-n-1})\otimes e_iu_2^n
 +e_iu_1^n\otimes \mu(e^*_i u_2^{-n-1})]
+[\mu(e_iu_1^{-n-1})\otimes e^*_iu_2^n +e^*_iu_1^n\otimes
\mu(e_iu_2^{-n-1})]\}=0.\eqno (2.35)$$

It is easy to know that
$\{\frac{e_i+e^*_i}{\sqrt{2}},\frac{e_i-e^*_i}{\sqrt{-2}}\}_{i\leq
K}$ is an orthonormal basis of $\frak g\oplus \frak g^*$ associated
to the bilinear form (2.33). Therefore by Theorem 2 with a direct
computation, we know that
$$
r=\sum_{i=1}^K[\frac{e_i\otimes e^*_i+e^*_i\otimes
e_i}{u_1-u_2}+\sum_{n=0}^L\mu(e^*_iu_1^{-n-1})\otimes e_iu_2^n
+\sum_{n=0}^L\mu(e_iu_1^{-n-1})\otimes e^*_iu_2^n]\eqno (2.36)$$ is
a rational solution of the CYBE  satisfying the unitary condition
for the classical double $\frak g\oplus\frak g^*$.

\section{Constructing a rational solution of the CYBE from an
$\mathcal{O}$-operator: coadjoint representations}

We have known that the construction of the rational solutions from
the ${\mathcal O}$-operators satisfying the equation (2.23) in
Theorem 2 partly depends on the existence of the Casimir element $t$
given by the nondegenerate symmetric invariant bilinear form, where
we use the key fact that $t\in \frak g\otimes \frak g$ is invariant
under the adjoint representation of a Lie algebra $\frak g$, that
is,
$$[x\otimes 1+1\otimes x, t]=0,\;\;\forall x\in \frak g.\eqno (3.1)$$
Actually, for any symmetric invariant tensor $t\in\frak g\otimes
\frak g$, it is easy to know that (\cite{CP})
$$r(u_1,u_2)=\frac{t}{u_1-u_2}\eqno (3.2)$$
satisfies the CYBE (1.1). In fact, it follows from
$$\sum_{i,j}[\frac{[a_i,a_j]\otimes b_i\otimes b_j}{(u_1-u_2)(u_1-u_3)}+\frac{a_i\otimes[b_i,a_j]\otimes b_j}{(u_1-u_2)(u_2-u_3)}
+\frac{a_i\otimes a_j\otimes[b_i,b_j]}{(u_1-u_3)(u_2-u_3)}]$$
$$=\sum_{i,j}[\frac{-a_i\otimes[b_i,a_j]\otimes b_j}{(u_1-u_2)(u_1-u_3)}+\frac{a_i\otimes[b_i,a_j]\otimes b_j}{(u_1-u_2)(u_2-u_3)}
+\frac{-a_i\otimes[b^i,a_j]\otimes b^j}{(u_1-u_3)(u_2-u_3)}]=0,\eqno
(3.3)$$ where $t=\sum\limits_{i} a_i\otimes
b_i=\sum\limits_{i}b_i\otimes a_i$. Note that here there are not any
constraint conditions for the Lie algebra $\frak g$ itself any more.
 Therefore it is natural to consider how to construct a rational solution of the CYBE with a form (2.4) from certain operators,
 where $t\in \frak g\otimes \frak g$ is symmetric invariant under the adjoint representation, as a generalization of the study in Section 2.

First we give some notations. Let $\frak g$ be a
(finite-dimensional) Lie algebra. Any $t\in \frak g\otimes \frak g$
can be regarded as a linear operator from $\frak g^*\rightarrow
\frak g$ by the following way
$$\langle t, a^*\otimes b^*\rangle=\langle t(a^*), b^*\rangle,\;\;\forall a^*,b^*\in \frak g^*.\eqno (3.4)$$
It can be defined from $\frak g^*[u,u^{-1}]$ to $\frak g[u,u^{-1}]$ by (it is still denoted by $t$)
$$t(a^*\otimes u^m)=t(a^*)\otimes u^m,\;\;\forall a^*\in \frak g^*, m\in \mathbb Z.\eqno (3.5)$$
On the other hand, let $\rho:\frak g\rightarrow gl(V)$ be a representation. The $V[u,u^{-1}]$ is still
a representation of $\frak g[u,u^{-1}]$ by (we still denote it by $\rho$)
$$\rho(x\otimes u^m)(v\otimes u^{n})=\rho(x)(v)\otimes u^{m+n},\;\;\forall x\in\frak g, v\in V, m,n\in \mathbb{Z}.\eqno (3.6)$$
Let ${\rm ad}$ be the adjoint representation of $\frak g$ and ${\rm ad}^*$ be the coadjoint representation
(the dual representation of the adjoint representation), that is,
$${\rm ad}(x)y=[x,y],\;\;\langle {\rm ad}^*(x) a^*, y\rangle=-\langle a^*,[x,y]\rangle,\;\;\forall x,y\in \frak g, a^*\in \frak g^*.\eqno (3.7)$$
In particular, if $t\in \frak g\otimes \frak g$ is symmetric
invariant under the adjoint representation, then
$$t({\rm ad}^*(x) a^*)=[x,t(a^*)],\;\;\forall x\in \frak g, a^*\in \frak g^*.\eqno (3.8)$$
In fact, let $t=\sum\limits_i a_i\otimes b_i$. Then for any $x\in \frak g, a^*,b^*\in \frak g^*$, we know that
\begin{eqnarray*}
\hspace{1cm}\langle t({\rm ad}^*(x) a^*), b^*\rangle
&=&\langle {\rm ad}^*(x) a^*\otimes b^*, t\rangle=\sum_i\langle -[x,a_i], a^*\rangle\langle b_i, b^*\rangle\\
&=& \langle -({\rm ad}(x) \otimes 1)t, a^*\otimes b^*\rangle
=\langle (1\otimes {\rm ad}(x)) t, a^*\otimes b^*\rangle\\
&=& -\langle a^*\otimes {\rm ad}^*(x) b^*, t\rangle=-\langle t(a^*), {\rm ad}^*(x) b^*\rangle
=\langle [x,t(a^*)], b^* \rangle.\hspace{1cm}(3.9)
\end{eqnarray*}
Moreover, since $t$ is symmetric, we have (the left hand side of the
equation (3.9))
$$\langle t({\rm ad}^*(x) a^*), b^*\rangle=\langle{\rm ad}^*(x)
a^*, t(b^*)\rangle=\langle -[x,t(b^*)], a^*\rangle=-\langle {\rm
ad}^*(t(b^*))a^*, x\rangle,\;\; \forall x\in \frak g, a^*,b^*\in
\frak g^*. \eqno (3.10)$$ By the equation (3.9), we know that
$${\rm ad}^*(t(a^*))b^*+{\rm ad}^*(t(b^*)a^*)=0,\;\;\;\forall
a^*,b^*\in \frak g^*.\eqno (3.11)$$

{\bf Theorem 3}\quad Let $\frak g$ be a  Lie algebra and $t\in \frak
g\otimes \frak g$ be symmetric invariant under the action of the
adjoint representation. Let $\{e_1,\cdots, e_K\}$ be a basis of
$\frak g$ and $\{e^*_1,\cdots,e^*_K\}$ be its dual basis. Then
$$r=\frac{t}{u_1-u_2}+\sum_{i=1}^K\sum_{n=0}^L T(e^*_iu_1^{-n-1})\otimes e_iu_2^n \eqno (3.12)$$
is a rational solution of the CYBE satisfying the unitary condition
(2.3) for $\frak g$ if the linear operator $T:\frak
g^*[u,u^{-1}]\rightarrow \frak g [u]\subset \frak g[u,u^{-1}]$
 satisfies the following conditions:
$$[T(f),T(g)]=T({\rm ad}^*(T(f))g-{\rm ad}^*(T(g))f-{\rm ad}^*(t(g))f),\;\;\forall f,g\in \frak g^*[u,u^{-1}];\eqno (3.13)$$
$$\sum_{i=1}^K\sum_{n=0}^L T(e^*_iu_1^{-n-1})\otimes e_iu_2^n+\sum_{i=1}^K\sum_{n=0}^L e_iu_1^n\otimes T(e^*_iu_2^{-n-1})=0;\eqno (3.14)$$
$$T(a^*u^p)=-t(a^*)u^p,\;\;\forall a^*\in \frak g^*; p\geq 0.\eqno (3.15)$$

In fact, let $t=t^{ij}e_i\otimes e_j\in \frak g\otimes \frak g$.
Then $t(e^*_i)=t^{ij}e_j$. Obviously, by the equations (3.8), (3.11)
and (3.15), for any $f\in\frak g[u]$ or $g\in \frak g[u]$, the
equation (3.13) holds automatically. With a similar study as in
Section 2, after substituting the equation (3.12) into the CYBE
(1.1), we can divide $[[r,r]]$ into four parts
$$[[r,r]]=\sum_{i,j=1}^K\sum_{p,q=0}^{2L}[T(e^*_iu_1^{-p-1}),T(e^*_ju_1^{-q-1})]\otimes e_iu_2^p\otimes e_ju_3^q+(A)+(B)+(C),\eqno (3.16)$$
where
\begin{eqnarray*}
\hspace{2cm}(A)&=&\sum_{i,j=1}^K\sum_{p,q=0}^{2L}T(e^*_iu_1^{-p-1})\otimes[e_iu_2^p,T(e^*_ju_2^{-q-1})]\otimes e_ju_3^q\\
 &&-\sum_{i,j=1}^K\sum_{p=0}^{2L}[\frac{T(e^*_iu_1^{-p-1})-T(e^*_iu_2^{-p-1})}{u_1-u_2},t(e^*_j)]\otimes
       e_j\otimes e_iu_3^p;\hspace{2.0cm} (3.17)\\
\hspace{2cm}(B)&=&\sum_{i,j=1}^K\sum_{p,q=0}^{2L}T(e^*_iu_1^{-p-1})\otimes T(e^*_ju_2^{-q-1})\otimes[e_iu_3^p,e_ju_3^q]\\
       &&+\sum_{i,j=1}^K\sum_{p=0}^{2L}[\frac{T(e^*_iu_1^{-p-1})-T(e^*_iu_3^{-p-1})}{u_1-u_3},t(e^*_j)]\otimes
       e_iu_2^p\otimes e_j;\hspace{2cm} (3.18)\\
\hspace{2cm}(C)&=&\sum_{i,j=1}^K\sum_{p=0}^{2L}T(e^*_iu_1^{-p-1})\otimes
       [e_i,t(e^*_j)]\otimes e_j\frac{u_2^p-u_3^p}{u_2-u_3}. \hspace{3.9cm} (3.19)
\end{eqnarray*}
Let $\mathcal B'$ be the bilinear form on the vector space $(\frak
g^*\oplus\frak g)[u,u^{-1}]$ given by the equation (2.28). By the
equation (3.14), we know that $\deg {\rm Im}T\leq L$, where ${\rm
Im}T$ is the image of the linear operator $T$. So we can set
$$T(e^*_iu^{-n-1})=\sum_{l=0}^L\alpha_l(e^*_iu^{-n-1})u^l,\eqno (3.20)$$
where $\alpha_l$ is a linear operator from $\frak g^*[u^{-1}]u^{-1}$
to
     $\frak g$, $l=0,1,\cdots, L$. Let $\alpha_l\equiv0$ when $l>L$. So
\begin{eqnarray*}
\hspace{0.3cm} (A)
&=&\sum_{i,j,k=1}^K\sum_{p,q,n=0}^{2L}T(e^*_iu_1^{-p-1})\otimes\mathcal
B' (e_iu_2^p,
 {\rm ad}^*(T(e^*_ju_2^{-q-1}))(e^*_ku_2^{-n-1})) e_ku_2^n\otimes e_ju_3^q\\
 &&+\sum_{i,j=1}^K\sum_{n,q=0}^{2L}[t(e^*_i)u_1^{-n-1},\sum_{l=n}^{2L-1}\alpha_{l+1}(e^*_ju_1^{-q-1})u_1^{l+1}]\otimes
 e_iu_2^n\otimes e_ju^q_3\\
 &=&\sum_{i,j,k=1}^K\sum_{p,q,n=0}^{2L}T(e^*_iu_1^{-p-1})\mathcal B'(e_iu_1^p,
 {\rm ad}^*(\sum_{l=0}^n\alpha_l(e^*_ju_1^{-q-1})u_1^{l})(e^*_ku_1^{-n-1}))\otimes e_ku_2^n\otimes e_ju_3^q\\
 &&+\sum_{i,j=1}^K\sum_{n,q=0}^{2L}[t(e^*_i)u_1^{-n-1},\sum_{l=n}^{2L-1}\alpha_{l+1}(e^*_ju_1^{-q-1})u_1^{l+1}]\otimes
 e_iu_2^n\otimes e_ju^q_3\\
 &=&\sum_{i,j=1}^K\sum_{n,q=0}^{2L}T({\rm ad}^*(\sum_{l=0}^n\alpha_l(e^*_ju_1^{-q-1})u_1^{l})(e^*_iu_1^{-n-1}))\otimes
 e_iu_2^n\otimes e_ju_3^q\\
 &&+\sum_{i,j=1}^K\sum_{n,q=0}^{2L}[t(e^*_i)u_1^{-n-1},\sum_{l=n}^{2L-1}\alpha_{l+1}(e^*_ju_1^{-q-1})u_1^{l+1}]\otimes
 e_iu_2^n\otimes e_ju^q_3\\
&=&\sum_{i,j=1}^K\sum_{n,q=0}^{2L}T({\rm
ad}^*T(e^*_ju_1^{-q-1})(e^*_iu_1^{-n-1}))\otimes
 e_iu_2^n\otimes e_ju^q_3.\hspace{4.2cm} (3.21)
\end{eqnarray*}
Note that in the last equation, we use the equations (3.8) and
(3.15). By the equation (3.14) and a similar study as above, we know
that
$$(B)=-\sum_{i,j=1}^K\sum_{n,q=0}^{2L}T({\rm ad}^*T(e^*_iu_1^{-q-1})(e^*_ju_1^{-n-1}))\otimes
 e_iu_2^q\otimes e_ju^n_3.\eqno (3.22)$$
Set $[e_i,e_j]=C_{ij}^ke_k$. Then
\begin{eqnarray*}
\hspace{2.5cm}(C)&=&\sum_{i=1}^K\sum_{p,q=0}^{2L}T(e^*_iu_1^{-p-q-2})t^{kj}C^l_{ik}\otimes
 e_lu_2^p\otimes e_ju^q_3\\
 &=&-\sum_{i,j=1}^K\sum_{p,q=0}^{2L}T({\rm ad}^*(t(e^*_iu_1^{-p-1}))e^*_ju_1^{-q-1})\otimes
 e_iu_2^p\otimes e_ju^q_3. \hspace{1.9cm} (3.23)
\end{eqnarray*}
Therefore $r$ given by the equation (3.12) is a solution of the CYBE
if the equations (3.13)-(3.15) hold. \hfill $\Box$

{\bf Example}\quad We give a concrete example of Theorem 3 as
follows. Let $\frak h$ be the 3-dimensional Heisenberg Lie algebra.
That is, there exists a basis $\{ e_1,e_2,e_3\}$ of $\frak h$
satisfying
$$[e_1,e_2]=e_3, \;[e_1,e_3]=[e_2,e_3]=0.\eqno (3.24)$$
Let $\{e_1^*,e_2^*,e_3^*\}$ be the dual basis. The coadjoint
representation ${\rm ad}^*$ is given as (only the non-zero actions
are given)
$${\rm ad}^*(e_1)e_3^*=-e_2^*,\;\;{\rm ad}^*(e_2)e_3^*=e_1^*.\eqno
(3.25)$$ Since $e_3$ is in the center of $\frak h$, $t=e_3\otimes
e_3$ is invariant under the action of the adjoint representation of
$\frak h$. Then
$$t(e_1^*)=0,\;t(e_2^*)=0, t(e_3^*)=e_3.\eqno (3.26)$$
So ${\rm ad}^*(t(a^*))b^*=0$ for any $a^*,b^*\in \frak h^*$.
Moreover, let $T_{\lambda_1,\lambda_2}:\frak
h^*[u,u^{-1}]\rightarrow \frak h[u]$ be a linear operator satisfying
(only the non-zero actions are given)
$$T(e^*_3u^{-2})=-(\lambda_1 e_1+\lambda_2e_2),
T(e^*_1u^{-1})=\lambda_1e_3u,T(e^*_2u^{-1})=\lambda_2e_3u;\;\;
T(e^*_3u^p)=-e_3u^p, \;\forall p\geq0,\eqno (3.27)$$ where
$\lambda_1,\lambda_2\in {\mathbb C}$. It is easy to know that
$T_{\lambda_1,\lambda_2}$ satisfies the equations (3.13)-(3.15). So
$$r(u_1,u_2)=\frac{e_3\otimes e_3}{u_1-u_2}+e_3\otimes (\lambda_1 e_1+\lambda_2e_2)u_1-(\lambda_1 e_1+\lambda_2e_2)\otimes e_3u_2\eqno (3.28)$$
is a rational solution of the CYBE satisfying the unitary condition
(2.3) for the Lie algebra $\frak h$. Although the solution (3.28)
seems a little trivial (all the commutators of $r$ are zero), it is
enough to illustrate the essential roles of the $\mathcal
O$-operators here. Moreover, it is easy to know that the above
construction can be generalized to any Lie algebra with a nonzero
center. \hfill $\Box$

Furthermore, in fact, the above construction can be regarded as a
natural generalization of Theorem 2 in the following sense. Let
$\frak g$ be a Lie algebra with a nondegenerate symmetric invariant
bilinear form ${\mathcal B}$. Let $\{e_1,\cdots, e_K\}$ be an
orthonormal basis of $\frak g$ and $\{e_1^*,\cdots, e_K^*\}$ be its
dual basis. Let $t=\sum\limits_i e_i\otimes e_i$ be the Casimir
element of $\frak g$. Then as a linear operator from $\frak g^*$ to
$\frak g$, $t$ satisfies $t(e_i^*)=e_i, i=1,\cdots, K$. Then as the
representations of $\frak g$, $\frak g^*$ can be identified with
$\frak g$ by the linear isomorphism $t$ in the following sense
$$\langle a^*, x\rangle=\mathcal B (t(a^*), x),
\;\;t({\rm ad}^*(x) a^*)=[x, t(a^*)],\;\;\forall x\in \frak g,
a^*\in\frak g^*.\eqno (3.29)$$ Therefore, we can get Theorem 2 from
Theorem 3 from the following correspondence:

\hspace{4cm} equation (3.14) $\Longleftrightarrow$ equation (2.24);

\hspace{4cm} equation (3.13) $\Longleftrightarrow$ equation (2.23);

\hspace{4cm} equation (3.15) $\Longleftrightarrow$ $\mu|_{{\frak
g}[u]}=-id$.

In particular, the equation (3.13) which is in fact well-defined on
$\frak g^*[u^{-1}]u^{-1}$ with a consistent extension (3.15) to
$\frak g^*[u]$ can be regarded as a generalization of the ${\mathcal
O}$-operator given by the equation (2.23) which is also well-defined
on $\frak g[u^{-1}]u^{-1}$ with a consistent extension $\mu|_{{\frak
g}[u]}=-id$ to $\frak g[u]$. Note that by the equation (3.11) or by
skew-symmetry of the Lie bracket (3.13),
$${\rm ad}^*(t(f))g+{\rm ad}^*(t(g))f=0,\;\;\forall f,g\in \frak
g^*[u,u^{-1}].\eqno (3.30)$$ Therefore, the above facts motivate us
to give a more general definition of an $\mathcal{O}$-operator which
is related to the rational solutions of the CYBE with the form (2.4)
(\cite{Ku}):

{\bf Definition}\quad Let $\frak g$ be a Lie algebra and
$\rho:\frak g\rightarrow gl(V)$ be a representation of  $\frak g$.
Suppose that there exists a skew-symmetric (bilinear) product $*$
on the vector space $V$ which gives a skew-symmetric bilinear
product on $V[u,u^{-1}]$ naturally by
$$(x\otimes u^{m})*(y\otimes u^{n})=(x*y)\otimes
u^{m+n},\;\;\forall x,y\in V, m,n\in {\mathbb Z}.\eqno (3.31)$$ A
linear map $T:V[u^{-1}]u^{-1}\rightarrow \frak g[u]\subset \frak
g[u,u^{-1}]$ with a suitable extension to $V[u]$ is called an
${\mathcal O}$-operator associated to $(\rho, V, *)$ if $T$
satisfies
$$[T(f),
T(g)]=T(\rho(T(f))g-\rho(T(g))f+f*g),\forall f,g\in
V[u,u^{-1}].\eqno (3.32)$$

Obviously, in the above sense, the equation (2.23) (with
$\mu|_{{\frak g}[u]}=-id$) in Theorems 1 and 2 gives an ${\mathcal
O}$-operator associated to the adjoint representation while the
equation (3.13) (with the equation (3.15)) in Theorem 3 gives an
${\mathcal O}$-operator associated to the coadjoint representation.

\section{Constructing a rational solution of the CYBE from an
$\mathcal{O}$-operator: general cases}

Not like the study on the construction of the constant solutions of
the CYBE from ${\mathcal O}$-operators in \cite{Bai}, it is not easy
to given an explicit construction of a rational solution of the CYBE
from an ${\mathcal O}$-operator for a general representation $(\rho,
V)$ and an arbitrary (skew-symmetric bilinear) product $*$ on $V$.
In this section, we consider the case that $(\rho, V)$ is still
arbitrary but the product $*$ on $V$ is trivial. Similar to the
study given in \cite{Bai}, the rational solutions (from the
following construction) of the CYBE from such $\mathcal O$-operators
are not for the Lie algebra $\frak g$ itself but for a larger Lie
algebra.

Let $\frak g$ still be a Lie algebra and $\rho: \frak g\rightarrow
gl(V)$ be a representation of $\frak g$. It is known that there is a
Lie algebra structure on a direct sum $\frak g\oplus V$ of the
underlying vector spaces $\frak g$ and $V$ given by
$$[e_1+x_1,e_2+x_2]=[e_1,e_2]+\rho(e_1)(x_2)-\rho(e_2)(x_1),\;\;\forall e_1,e_2\in \frak g, x_1,x_2\in V.\eqno (4.1)$$
It is denoted by $\frak g\ltimes_\rho V$. On the other hand, let
$\rho^*:\frak g\rightarrow gl(V^*)$ be the dual representation of
$(\rho, V)$ of the Lie algebra $\frak g$, that is,
$$\langle \rho^*(e)x^*, y\rangle=-\langle x^*, \rho(e) y\rangle,\;\;\forall e\in \frak g, x^*\in V^*, y\in V.
\eqno (4.2)$$ Then both $V$ and $V^*$ can be the representations of
the Lie algebra $\frak g\ltimes_{\rho^*} V^*$ by the zero-extension,
that is (we still denote them by $\rho$ and $\rho^*$ respectively),
$$\rho(e+x^*) y=\rho (e) y;\;\; \rho^*(e+x^*)z^*=\rho^*(e)
z^*,\;\;\forall e\in \frak g, x^*,z^*\in V^*, y\in V.\eqno (4.3)$$
Moreover, by the equation (3.6), both $V[u,u^{-1}]$ and
$V^*[u,u^{-1}]$ are the representations of the Lie algebra $\frak
g\ltimes_{\rho^*}V^*[u,u^{-1}]$.

{\bf Theorem 4}\quad Let $\frak g$ be a Lie algebra and $\rho: \frak
g\rightarrow gl(V)$ be a representation of $\frak g$. Let $t\in
V^*\otimes V^*$ be symmetric invariant under the action of the dual
representation $\rho^*$. Let $\{w_1,\cdots, w_N\}$ be a basis of $V$
and $\{w_1^*,\cdots,w_N^*\}$ be its dual basis. Then
$$r=\frac{2t}{u_1-u_2}+\sum_{i=1}^N\sum_{k=0}^L
T(w_iu_1^{-k-1})\otimes w^*_iu_2^k-\sum_{i=1}^N\sum_{k=0}^L
w^*_iu_1^k\otimes T(w_iu_2^{-k-1})\eqno (4.4)$$ is a rational
solution of the CYBE satisfying the unitary condition (2.3) for the
Lie algebra $\frak g\ltimes_{\rho^*} V^*$ if the linear operator $T:
V[u,u^{-1}]\rightarrow (\frak g\ltimes_{\rho^*} V^* )[u]$ with $\deg
{\rm Im} T\leq L$ satisfies the following conditions:
$$[T(f),
T(g)]=T(\rho(T(f))g-\rho(T(g))f),\forall f,g\in V[u,u^{-1}];\eqno
(4.5)$$
$$T(V[u^{-1}]u^{-1})\subset \frak g[u],\;\;T(wu^p)=-t(w)u^p,\;\;\forall w\in V, p\geq 0.\eqno (4.6)$$
In fact, let $t=t^{ij}w^*_i\otimes w^*_j\in V^*\otimes V^*$. Then
$t(w_i)=t^{ij}w^*_j$. Moreover, since $t$ is symmetric invariant
under the action of the dual representation $\rho^*$, we know that
$$t(\rho(e)w)=\rho^*(e)t(w),\;\;\forall e\in \frak g, w\in V,\eqno
(4.7)$$ by replacing ${\rm ad}^*$ by $\rho$ in the equation (3.9).
Obviously, by the equations (4.6) and (4.7), for any $f\in V[u]$ or
$g\in V[u]$, the equation (4.5) holds automatically. Let $\mathcal
B'$ be the bilinear form on the vector space $(V^*\oplus
V)[u,u^{-1}]$ given by the equation (2.28). On the other hand, as in
the above section, from $\deg {\rm Im}T\leq L$, we can set
$$T(w_iu^{-k-1})=\sum_{l=0}^L\alpha_l(w_iu^{-k-1})u^l,\eqno (4.8)$$
where $\alpha_l$ is a linear operator from $ V[u^{-1}]u^{-1}$ to
$\frak g$, $l=0,1,\cdots, L$. Let $\alpha_l\equiv0$ when $l>L$.

Substituting the equation (4.4) into the CYBE (1.1), we know that
$$[[r,r]]=(A)+(B)+(C),\eqno (4.9)$$
where
\begin{eqnarray*}
\hspace{0.5cm}(A)&=&\sum_{i,j=1}^N\sum_{n,q=0}^{2L}[T(w_iu_1^{-n-1}),T(w_ju_1^{-q-1})]\otimes
w^*_iu_2^n\otimes w^*_ju_3^q\\
&&+\sum_{i,j=1}^N\sum_{p,q=0}^{2L}T(w_iu_1^{-p-1})\otimes[w^*_iu_2^p,T(w_ju_2^{-q-1})]\otimes w^*_ju_3^q\\
 &&-\sum_{i,j=1}^N\sum_{p=0}^{2L}[\frac{T(w_iu_1^{-p-1})-T(w_iu_2^{-p-1})}{u_1-u_2},t(w_j)]\otimes
       w^*_j\otimes w^*_iu_3^p\\
       &&-\sum_{i,j=1}^N\sum_{p,q=0}^{2L}T(w_iu_1^{-p-1})\otimes w^*_ju_2^q\otimes[w^*_iu_3^p,T(w_ju_3^{-q-1})]\\
       &&+\sum_{i,j=1}^N\sum_{p=0}^{2L}[\frac{T(w_iu_1^{-p-1})-T(w_iu_3^{-p-1})}{u_1-u_3},t(w_j)]\otimes
       w^*_iu_2^p\otimes w^*_j;\hspace{3.0cm} (4.10)\\
(B)&=&\sum_{i,j=1}^N\sum_{p,q=0}^{2L}\rho^*(T(w_ju_1^{-p-1}))(w^*_iu_1^q)\otimes
       T(w_iu_2^{-q-1})\otimes w^*_ju_3^p\\
       &&-\sum_{i,j=1}^N\sum_{p,q=0}^{2L}w^*_iu_1^p\otimes
               [T(w_iu_2^{-p-1}),T(w_ju_2^{-q-1})]\otimes
               w^*_ju_3^q\\
               &&-\sum_{i,j=1}^N\sum_{p,q=0}^{2L}w^*_iu_1^p\otimes
 T(w_ju_2^{-q-1})\otimes\rho^*(T(w_iu_3^{-q-1}))(w^*_ju_3^p)\\
  \end{eqnarray*}
\begin{eqnarray*}
 &&-\sum_{i,j=1}^N\sum_{p=0}^{2L}[\frac{T(w_iu_1^{-p-1})-T(w_iu_2^{-p-1})}{u_1-u_2},t(w_j)]\otimes
       w^*_j\otimes w^*_iu_3^p\\
       &&-\sum_{i,j=1}^N\sum_{p=0}^{2L}\frac{w^*_iu_1^p\otimes[T(w_iu_2^{-p-1})-T(w_iu_3^{-p-1}),t(w_j)]\otimes
               w^*_j}{u_2-u_3};\hspace{3.0cm} (4.11)\\
(C)&=&-\sum_{i,j=1}^N\sum_{p,q=0}^{2L}\rho^*(T(w_iu_1^{-p-1}))(w^*_ju_1^q)\otimes
       w^*_iu_2^p\otimes T(w_ju_3^{-q-1})\\
       &&+\sum_{i,j=1}^N\sum_{p,q=0}^{2L}w^*_iu_1^p\otimes
               \rho^*(T(w_iu_2^{-q-1}))(w^*_ju_2^p)\otimes
               T(w_ju_3^{-q-1})\\
               &&+\sum_{i,j=1}^N\sum_{p,q=0}^{2L}w^*_iu_1^p\otimes
 w^*_ju_2^q\otimes[T(w_iu_3^{-p-1}),T(w_ju_3^{-q-1})]\\
 &&+\sum_{i,j=1}^N\sum_{p=0}^{2L}[\frac{T(w_iu_1^{-p-1})-T(w_iu_3^{-p-1})}{u_1-u_3},t(w_j)]\otimes
       w^*_iu_2^p\otimes w^*_j\\
       &&-\sum_{i,j=1}^N\sum_{p=0}^{2L}\frac{w^*_iu_1^p\otimes[T(w_iu_2^{-p-1})-T(w_iu_3^{-p-1}),t(w_j)]\otimes
               w^*_j}{u_2-u_3}.\hspace{3.0cm} (4.12)
\end{eqnarray*}
We can divide $(A)$ given by the equation (4.10) into three parts:
$$(A)=\sum_{i,j=1}^N\sum_{n,q=0}^{2L}[T(w_iu_1^{-n-1}),T(w_ju_1^{-q-1})]\otimes
w^*_iu_2^n\otimes w^*_ju_3^q +(A_1)+(A_2),\eqno (4.13)$$ where
\begin{eqnarray*}
\hspace{1.5cm}(A_1)&=&\sum_{i,j=1}^N\sum_{p,q=0}^{2L}T(w_iu_1^{-p-1})\otimes[w^*_iu_2^p,T(w_ju_2^{-q-1})]\otimes w^*_ju_3^q\\
 &&-\sum_{i,j=1}^N\sum_{p=0}^{2L}[\frac{T(w_iu_1^{-p-1})-T(w_iu_2^{-p-1})}{u_1-u_2},t(w_j)]\otimes
       w^*_j\otimes w^*_iu_3^p;\hspace{1.9cm} (4.14)\\
\hspace{1.5cm}(A_2)&=&-\sum_{i,j=1}^N\sum_{p,q=0}^{2L}T(w_iu_1^{-p-1})\otimes w^*_ju_2^q\otimes[w^*_iu_3^p,T(w_ju_3^{-q-1})]\\
       &&+\sum_{i,j=1}^N\sum_{p=0}^{2L}[\frac{T(w_iu_1^{-p-1})-T(w_iu_3^{-p-1})}{u_1-u_3},t(w_j)]\otimes
       w^*_iu_2^p\otimes w^*_j.\hspace{1.9cm} (4.15)
\end{eqnarray*}
Moreover,
\begin{eqnarray*}
\hspace{0.1cm}(A_1)
&=&\sum_{i,j,k=1}^N\sum_{p,q,n=0}^{2L}T(w_iu_1^{-p-1})\mathcal B'(
w^*_iu_1^p,
 \rho(\sum_{l=0}^n\alpha_l(w_ju_1^{-q-1})u_1^{l})(w_ku_1^{-n-1}))\otimes w^*_ku_2^n\otimes w^*_ju_3^q\\
&&+\sum_{i,j=1}^N\sum_{n,q=0}^{2L}[t(w_i)u_1^{-n-1},\sum_{l=n}^{2L-1}\alpha_{l+1}(w_ju_1^{-q-1})u_1^{l+1}]\otimes
 w^*_iu_2^n\otimes w^*_ju^q_3\\
 \end{eqnarray*}
\begin{eqnarray*}
 &=&\sum_{i,j=1}^N\sum_{n,q=0}^{2L}T(\rho(\sum_{l=0}^n\alpha_l(w_ju_1^{-q-1})u_1^{l})(w_iu_1^{-n-1}))\otimes
 w^*_iu_2^n\otimes w^*_ju_3^q\\
\hspace{0.7cm}
&&+\sum_{i,j=1}^N\sum_{n,q=0}^{2L}[t(w_i)u_1^{-n-1},\sum_{l=n}^{2L-1}\alpha_{l+1}(w_ju_1^{-q-1})u_1^{l+1}]\otimes
 w^*_iu_2^n\otimes w^*_ju^q_3\\
&=&\sum_{i,j=1}^N\sum_{n,q=0}^{2L}T(\rho(
T(w_ju_1^{-q-1}))(w_iu_1^{-n-1}))\otimes
 w^*_iu_2^n\otimes w^*_ju^q_3.\hspace{4cm} (4.16)
\end{eqnarray*}
Note that in the last equation, we use the equations (4.6) and
(4.7). Similarly, we have
$$(A_2)=-\sum_{i,j=1}^N\sum_{n,q=0}^{2L}T(\rho (T(w_iu_1^{-q-1}))(w_ju_1^{-n-1}))\otimes
 w^*_iu_2^q\otimes w^*_ju^n_3.\eqno (4.17)$$
Therefore
\begin{eqnarray*}
\hspace{0.2cm}(A)
&=&\sum_{i,j=1}^N\sum_{n,q=0}^{2L}\{[T(w_iu_1^{-n-1}),T(w_ju_1^{-q-1})]
+T(\rho( T(w_ju_1^{-q-1}))(w_iu_1^{-n-1}))\\
&&-T(\rho( T(w_iu_1^{-n-1}))(w_ju_1^{-q-1}))\}\otimes
 w^*_iu_2^n\otimes w^*_ju^q_3. \hspace{5cm} (4.18)
 \end{eqnarray*}
With a similar discussion as above, we know that
\begin{eqnarray*}
\hspace{0.7cm}[[r,r]]&=&\sum_{i,j=1}^N\sum_{n,q=0}^{2L}\{[T(w_iu_1^{-n-1}),T(w_ju_1^{-q-1})]+T(\rho(
T(w_ju_1^{-q-1}))(w_iu_1^{-n-1}))\\ &&-T(\rho(
T(w_iu_1^{-n-1}))(w_ju_1^{-q-1}))\}\otimes
 w^*_iu_2^n\otimes w^*_ju^q_3\\
&&-\sum_{i,j=1}^N\sum_{n,q=0}^{2L}w^*_iu_1^n\otimes
               \{[T(w_iu_2^{-n-1}),T(w_ju_2^{-q-1})]+T(\rho(
T(w_ju_2^{-q-1}))(w_iu_2^{-n-1}))\\ &&-T(\rho(
T(w_iu_2^{-n-1}))(w_ju_2^{-q-1}))\}\otimes
               w^*_ju_3^q\\
&&+\sum_{i,j=1}^N\sum_{n,q=0}^{2L}w^*_iu_1^n\otimes
 w^*_ju_2^q\otimes\{[T(w_iu_3^{-n-1}),T(w_ju_3^{-q-1})]\\
 &&+T(\rho(
T(w_ju_3^{-q-1}))(w_iu_3^{-n-1}))-T(\rho(
T(w_iu_3^{-n-1}))(w_ju_3^{-q-1}))\}. \hspace{1.7cm} (4.19)
\end{eqnarray*}
Therefore $r$ given by the equation (4.4) is a solution of the CYBE
if the equations (4.5)-(4.6) hold. Obviously, $r$ satisfies the
unitary condition (2.3).\hfill $\Box$

Note that in the equation (4.4), $t$ is taken in the vector space
$V^*\otimes V^*\subset (\frak g\ltimes_{\rho^*} V^*)\otimes (\frak
g\ltimes_{\rho^*} V^*)$ which there is not any part in $\frak g$.
Otherwise, it would involve the actions between the Lie algebra
$\frak g$ itself which might be  very complicated and a little far
away from the equation (4.5) defining the ${\mathcal O}$-operator
associated to any arbitrary representation (in fact, it might
involve the coadjoint representation as given in Section 3).

At the end of this section, we consider two special cases and then
compare them with the relative study in Section 2 and Section 3
respectively.

(Case I)\quad The representation $\rho$ is taken as the adjoint
representation ${\rm ad}$. Then by Theorem 4, we can get a rational
solution of the CYBE with the form (4.4) for the Lie algebra ${\frak
g}\ltimes_{{\rm ad}^*}\frak g^*$, where $t\in \frak g^*\otimes \frak
g^*$ and the linear operator $T$ is from $\frak g[u,u^{-1}]$ to
${\frak g}\ltimes_{{\rm ad}^*}\frak g^*[u]$. On the other hand, it
is known (\cite{CP}) that ${\frak g}\ltimes_{{\rm ad}^*}\frak g^*$
is the classical double of the trivial Lie bialgebra structure (that
is, $\delta_{\frak g}=0$) on the Lie algebra $\frak g$. Therefore by
the study at the end of section 2, there is another (completely
different) rational solution of the CYBE with the form (2.31) for
the same Lie algebra ${\frak g}\ltimes_{{\rm ad}^*}\frak g^*$, where
$t\in ({\frak g}\ltimes_{{\rm ad}^*}\frak g^*)\otimes ({\frak
g}\ltimes_{{\rm ad}^*}\frak g^*)$ and $T=\mu$ is a linear operator
from ${\frak g}\ltimes_{{\rm ad}^*}\frak g^*[u,u^{-1}]$ to ${\frak
g}\ltimes_{{\rm ad}^*}\frak g^*[u]$.

(Case II)\quad The representation $\rho$ is taken as the coadjoint
representation ${\rm ad}^*$. Then by Theorem 4, we can get a
rational solution of the CYBE with the form (4.4) for the Lie
algebra ${\frak g}\ltimes_{\rm ad}\frak g$, where $t\in \frak
g\otimes \frak g$ and the linear operator $T$ is from $\frak
g^*[u,u^{-1}]$ to ${\frak g}\ltimes_{\rm ad}\frak g[u]$.  In
particular, $r(u_1,u_2)$ given by the equation (4.4) in this case is
in the vector space $\frak g[u_1]\otimes \frak g[u_2]$. Despite it,
this $r(u_1,u_2)$ is not a rational solution of the CYBE for the Lie
algebra $\frak g$ itself which only involves the adjoint action in
general. In fact, although the forms of $t$ and
$T(w_iu_1^{-k-1})\otimes w^*_iu_2^k$ involve the vector space $\frak
g\otimes \frak g$, they actually involve the vector space $(0,\frak
g)\otimes (0,\frak g)$ and $(\frak g,0)\otimes (0,\frak g)$
respectively. However, the commutators (zero) between $(0, \frak g)$
and $(0,\frak g)$ are different with the commutators (adjoint
action) between $(\frak g,0)$ and $(\frak g, 0)$ or $(\frak g,0)$
and $(0,\frak g)$. Nevertheless, by the proof of Theorems 3 and 4,
we can prove that if $T:\frak g^*[u,u^{-1}]\rightarrow \frak g[u]$
satisfies equation (3.15), the unitary condition (3.14) and
$$[T(f),T(g)]=T({\rm ad}^*(T(f))g-{\rm ad}^*(T(g))f),\;\;\forall f,g\in \frak g^*[u^{-1},u],\eqno (4.20)$$
 and
$${\rm ad}^*(t(g))f=0,\;\;\forall f,g\in \frak g^*[u^{-1},u],\eqno (4.21)$$
then
$$r=\frac{2t}{u_1-u_2}+2\sum_{i=1}^K\sum_{k=0}^L T(e^*_iu_1^{-k-1})\otimes e_iu_2^k ,\eqno (4.22)$$
is a rational solution of the CYBE for the Lie algebra $\frak g$,
which coincides with the construction from Theorem 3 under the
condition (4.21) since $r$ is a solution of the CYBE if and only if
$2r$ is a solution of the CYBE.

\section{Conclusions and discussion}

From the study in the previous sections, we give the following
conclusions and discussion.

(1) The rational solutions of the CYBE for the complex simple Lie
algebras are interpreted in terms of the ${\mathcal O}$-operators
(associated to the adjoint representations). Furthermore, the
${\mathcal O}$-operators (associated to the suitable
representations) can be used to construct explicitly the rational
solutions for a Lie algebras with a nondegenerate symmetric
invariant bilinear form (adjoint representation), a Lie algebra with
a symmetric invariant tensor under the action of the adjoint
representation (coadjoint representation), a semidirect sum of a Lie
algebra and the dual representation of its representation with a
symmetric invariant tensor under the action of the dual
representation (an arbitrary representation). All of the solutions
have a uniform form as
$$r(u_1,u_2)=\frac{t}{u_1-u_2}+r_0(u_1,u_2),\eqno (5.1)$$
where $t$ is the symmetric invariant tensor and $r_0$ is a
polynomial defined by an ${\mathcal O}$-operator. We call the
equation (5.1) a Drinfeld form (\cite{D1},\cite{D2}). Note that in
the above construction the existence of a symmetric invariant tensor
$t$ is necessary and it plays an essential role in the concrete
definition of an ${\mathcal O}$-operator defining the polynomial
$r_0$.

(2) Comparing with the study of the operator approach to the
constant CYBE (\cite{Bai}), we find that, roughly speaking, there
are the following ``correspondence":
\begin{eqnarray*}
{\rm adjoint}\; {\rm representation:}\; {\rm equation}\; (2.23) &
\longleftrightarrow& {\rm equation}\; (1.5)\; (\mbox{{\rm Semonov-Tian-Shansky)}}\\
{\rm coadjoint}\; {\rm representation:}\; {\rm equation}\; (3.13) &
\longleftrightarrow & {\rm equation}\; (1.6)\; {\rm
(Kupershmidt)}\\
{\rm an}\; {\rm arbitrary}\;{\rm  representation:}\; {\rm
equation}\; (4.5) &\longleftrightarrow& {\rm equation}\; (1.7)\;
({\rm Kupershmidt}\;{\rm and}\; [14])
\end{eqnarray*}
However, the ``correspondence" is in a ``rough" level, since not
like in the study of the constant CYBE that the construction from
the ${\mathcal O}$-operators by the equations (1.5) and (1.6)  can
be obtained through the uniform construction from ${\mathcal
O}$-operators by the equation (1.7) as the special cases
(\cite{Bai}), the construction from ${\mathcal O}$-operators by the
equations (2.23) and (3.13) are usually quite different with the
construction from ${\mathcal O}$-operators by the equation (4.5) in
the corresponding cases (see the discussion at the end of section
4). In fact, it is due to the inhomogeneous term $T(f*g)$ appearing
in the definition (3.32) which is closely related to the symmetric
invariant tensor $t$.

(3) The form of the equation (2.23) with the inhomogeneous term
$\mu[f,g]$ which gives an ${\mathcal O}$-operator associated to the
adjoint representation corresponds precisely to the linear operator
$R':\frak g\rightarrow \frak g$ in the constant case satisfying
$$[R'(x), R'(y)]=R'([R'(x),y]+[x,R'(y)])+R'[x,y],\;\;\forall x,y\in \frak g.\eqno (5.2)$$
The above equation is exactly the Rota-Baxter relation of weight
$-1$ in the version of Lie algebras (\cite{E}, \cite{EGK}). By
letting $R'=1-2R$, the equation (5.2) is equivalent to the operator
form (\cite{STS2})
$$[R(x), R(y)]=R([R(x),y]+[x,R(y)])-[x,y],\;\;\forall x,y\in \frak g\eqno (5.3)$$
of the (constant) modified classical Yang-Baxter equation (MCYBE,
\cite{CP}) satisfying
$$[[r,r]]\;\; {\rm is}\;\;{\rm invariant}\;\;{\rm under}\;\;{\rm
the}\;\;{\rm adjoint}\;\;{\rm action}.\eqno (5.4)$$ Moreover, the
product $[\;,\;]_1$ given by
$$[x,y]_1=[R'(x),y]+[x,R'(y)]+[x,y],\;\;\forall x,y\in \frak g\eqno (5.5)$$
defines a Lie algebra and $R'$ is a homomorphism between two Lie
algebras. On the other hand, for the equation (2.23), if we define
$$f\circ g=[\mu (f), g]+1/2 [f,g],\;\;\forall f,g\in \frak
g[u,u^{-1}],\eqno (5.6)$$ then $(\frak g[u,u^{-1}],\circ)$ is a
Lie-admissible algebra satisfying that
$$[f,g]_1=f\circ g-g\circ f=[\mu (f),g]+\mu[f,\mu (g)]+[f,g],\;\;\forall f,g\in \frak
g[u,u^{-1}],\eqno (5.7)$$ defines a Lie algebra and $\mu$ is a
homomorphism of Lie algebras.

(4) As we have already pointed out in Section 2, besides the complex
simple Lie algebras, the construction from ${\mathcal O}$-operators
in this paper may not give all the rational solutions of the CYBE.
What we have done in this paper is just an effort to provide certain
helpful ideas to construct the solutions of the CYBE with parameters
for the general Lie algebras which is still an open question.

(5) It is natural to consider the corresponding Lie bialgebra
structures from the rational solutions (5.1) of the CYBE
constructing from the $\mathcal O$-operators through
$$\delta(f)(u,v)=[f(u)\otimes 1+1\otimes f(v), r(u,v)],\;\;\forall f\in \frak
g[u,u^{-1}].\eqno (5.8)$$ It is also natural and important to
consider the quantization of these Lie bialgebra structures.

\section*{Acknowledgments}

This work was supported in part by  the National Natural Science
Foundation of China (10571091, 10621101), NKBRPC (2006CB805905),
SRFDP (200800550015), Program for New Century Excellent Talents in
University. Part of this work was done while the second author was
visiting Korea Institute for Advanced Study (KIAS). He would like to
thank KIAS and his host Dr. Dafeng Zuo for the hospitality and the
valuable discussions.

\bibliography{}

\end{document}